\documentclass[prl,twocolumn,showpacs]{revtex4}
\usepackage{graphicx}
\begin{document}

\title{Nondestructive readout for a superconducting flux qubit}

\author{A. Lupa\c scu}
\affiliation{Kavli Institute of Nanoscience, Delft University of
Technology, PO Box 5046, 2600 GA Delft, The Netherlands}
\author{C. J. M. Verwijs}
\affiliation{Kavli Institute of Nanoscience, Delft University of
Technology, PO Box 5046, 2600 GA Delft, The Netherlands}
\author{R. N. Schouten}
\affiliation{Kavli Institute of Nanoscience, Delft University of
Technology, PO Box 5046, 2600 GA Delft, The Netherlands}
\author{C. J. P. M. Harmans}
\affiliation{Kavli Institute of Nanoscience, Delft University of
Technology, PO Box 5046, 2600 GA Delft, The Netherlands}
\author{J. E. Mooij}
\affiliation{Kavli Institute of Nanoscience, Delft University of
Technology, PO Box 5046, 2600 GA Delft, The Netherlands}

\date{\today}

\begin{abstract}

We present a new readout method for a superconducting flux qubit, based on the measurement of the Josephson inductance of a superconducting quantum interference device that is inductively coupled to the qubit. The intrinsic flux detection efficiency and back-action are suitable for a fast and nondestructive determination of the quantum state of the qubit, as needed for readout of multiple qubits in a quantum computer. We performed spectroscopy of a flux qubit and we measured relaxation times of the order of 80 $\mu s$.
\end{abstract}
\maketitle

Suitably designed superconducting circuits, based on Josephson junctions, behave as quantum two level systems. Due to scalability and flexibility in their design parameters, they are promising candidates for quantum bits (or qubits), which are the basic units in a quantum information processor~\cite{nielsen_2000_1}. In such circuits, coherent evolution for single qubits was observed
~\cite{nakamura_1999_1,vion_2002_1,yu_2002_1,martinis_2002_1,chiorescu_2003_1,duty_2004_1},
and a conditional gate for two qubits was demonstrated~\cite{yamamoto_2003_1}.

Flux qubits consist of a superconducting loop interrupted by one or more Josephson junctions. The basis states have opposite persistent current. The quantum state can be read out by measuring the generated magnetic flux, using a DC superconducting quantum interference device (DC-SQUID). The critical current of the SQUID depends on this flux and is usually measured by determining the maximum supercurrent, where the device switches to a finite voltage. When the SQUID is in this voltage state it generates quasiparticles that later recombine with a burst of energy. It also radiates strong high frequency signals into the whole circuit with, in future, multiple qubits and readout devices. Inevitably, significant quantum information is destroyed apart from the consequence of reading out one qubit. In contrast, we now measured the SQUID critical current by determining the Josephson inductance, without dissipation in the SQUID system. In this Letter, we discuss the intrinsic properties of this inductive readout and we present the first results of measurements on a flux qubit.

\begin{figure}[!]
\includegraphics[width=3.4in]{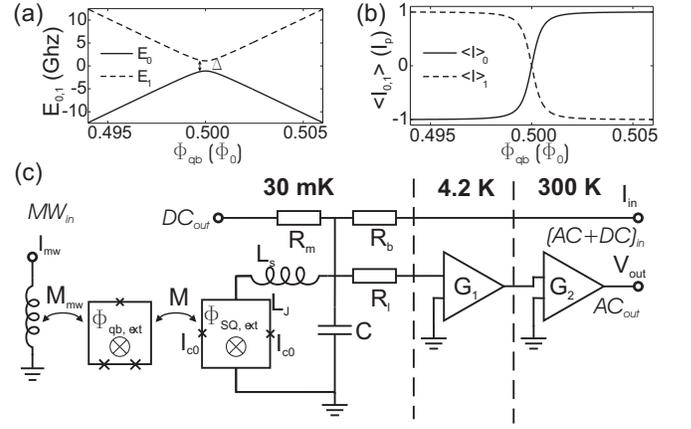}
\caption{\label{fig1} (a) Ground and excited state energy levels versus the magnetic flux in the PCQ loop, for $E_{J}/E_{C} = 110$, $E_{C}/h = 4.6$ GHz and ${\alpha = 0.75}$. (b) Expectation value of the qubit persistent current in the ground and excited state. (c) Diagram of the experimental setup. A waveguide is used to apply a small magnetic flux with amplitude $M_{mw}I_{mw}$ in the qubit loop. A coaxial line is used to apply the AC and DC bias current. The output AC voltage is amplified at 4.2 K and room temperature with gains ${G_{1} = 14}$ dB and ${G_{2} = 70}$ dB. The values of the components are ${L_{s} = 2.35}$ nH, ${C = 12}$ pF, ${R_{b} = 5.6}$ k$\Omega$, ${R_{m} = 11}$ k$\Omega$ and ${R_{l} = 820}$ $\Omega$.}
\end{figure}

In the experiments, we use a persistent current qubit (PCQ)~\cite{mooij_1999_1}. The PCQ is a flux qubit consisting of a small inductance superconducting loop interrupted by three Josephson junctions. Two of the three junctions are equal, characterized by the Josephson coupling energy $E_{J}$ and the charging energy $E_{C}$, and the third one is smaller by a factor $\alpha$. At low temperatures and appropriate values of $E_{J}>E_{C}$ and $\alpha$, and with an external magnetic flux $\Phi_{qb}$ close to $(2n+1)\Phi_{0}/2$ (n=integer) in the loop, the circuit behaves as a two level system. In the basis of two circulating current or flux states, the Hamiltonian reads
\begin{equation}\label{eq_Ham}
  H={\textstyle\frac{1}{2}}(\epsilon\sigma_{\!z} + \Delta\sigma_{\!x}) \;,
\end{equation}
in which $\sigma_{i} (i=x,y,z)$ are the Pauli matrices and $\epsilon=2I_{p}(\Phi_{qb}-(2n+1)\Phi_{0}/2)$. The minimum energy level splitting $\Delta$ and the maximum qubit persistent current $I_{p}$ are determined by $E_{J}$, $E_{C}$ and $\alpha$. The two energy eigenstates (which are linear superpositions of the current states) each have an expectation value of the circulating current given by the derivative of the corresponding energy eigenvalue with respect to $\Phi_{qb}$ (see Fig. \ref{fig1}a and b).

A DC-SQUID magnetometer contains a loop interrupted by two Josephson junctions. The SQUID critical current is $I_{c}=2I_{c0}|\cos(\pi\Phi_{SQ}/\Phi_{0})|$, where $I_{c0}$ is the critical current of a single junction and $\Phi_{SQ}$ is the flux in the SQUID loop, which contains a contribution due to the coupled flux qubit. The voltage V and the supercurrent I of the SQUID are related to the difference in superconducting phase across the terminals of the DC-SQUID, $\gamma$, through the Josephson relations: $V=\phi_{0}\dot{\gamma}$ and $I=I_{c}\sin(\gamma)$, in which $\phi_{0}=\Phi_{0}/2\pi$. For small variations of the current around an average value $I_{0}<I_{c}$, the DC-SQUID can be described as an inductor, with the Josephson inductance $L_{J}=\phi_{0}/\sqrt{I_{c}^2-I_{0}^2}$.

The Josephson inductance is measured by injecting an AC current $I_{AC}$ of frequency $\nu$, generating a voltage $\sim(I_{AC}/I_{c})\Phi_{0}\nu$. The injected current leads to a small magnetic flux in the qubit loop with significant frequency components at 0, $\nu$ and $2\nu$. To avoid qubit excitation $\nu$ is taken well below the qubit level splitting. The qubit induced change in the inductance is small ($\sim 1\%$ in our experiment). To enhance the measurement sensitivity, a resonant circuit is formed by adding a shunt capacitor to the SQUID. If the frequency $\nu$ is close to the resonance frequency of this circuit, a small change in the Josephson inductance leads to a large change in the AC voltage, resulting in efficient state detection. Our method is similar to the RF - single electron transistor, used to measure charge qubits~\cite{duty_2004_1}, but avoids the measurement of a dissipative circuit element. A readout scheme similar to ours was proposed for charge qubits~\cite{zorin_2001_1}, and a detection scheme exploiting the non-linearity of $I(\gamma)$ was shown in~\cite{siddiqi_2003_2}. 
A PCQ was studied in~\cite{grajcar_2004_1} by measuring the qubit loop susceptibility using a coupled linear resonant circuit, with response time on microsecond time scales.

Figure 1c shows the schematic diagram of the circuit comprising the qubit, DC-SQUID and microwave excitation line. 
The combined PCQ and SQUID system is fabricated on a single oxidized silicon substrate using electron beam lithography and double-angle shadow evaporation of aluminum films. The design values for the qubit Josephson junctions are \(E_{c}/h = 4.6\) GHz, \(E_{J}/E_{C} = 110\) and \(\alpha = 0.75\), which yields $I_{p} = 660$ nA. The areas of the PCQ and SQUID are, respectively, ${A_{qb} = 100}$ $\mu$m$^2$ and ${A_{SQ} = 128}$ $\mu$m$^2$. The chip containing the qubit and SQUID is attached to a printed circuit board on which the capacitor $C$ and the resistors $R_{l}$, $R_{b}$ and $R_{m}$ are mounted. Capacitor $C$ is positioned close to the SQUID to reduce the stray inductance $L_s$. Resistor $R_b$ determines the (AC and DC) driving current to the SQUID resonator. Applying a DC current and measuring the SQUID DC voltage via resistor $R_m$ allows to determine the switching current. Applying an AC current to the SQUID resonator generates an AC voltage, which is coupled via the resistor $R_{l} = 820$ $\Omega$ to the $50$ $\Omega$ input impedance of the low noise 0.5-1 GHz amplifier. With $R_{m},R_{b}>>R_{l}>>50$ $\Omega$, the quality factor $Q\simeq40$ is mostly determined by the coupling resistor $R_{l}$. An external superconducting coil generates the magnetic fluxes $\Phi_{qb, ext}$, in the qubit loop and $\Phi_{SQ,ext}$, in the SQUID loop. Qubit excitation is performed by a microwave magnetic flux, using a line with mutual inductance $M_{mw}$=0.12 pH to the qubit. Appropriate noise filtering is employed in all signal lines. The printed circuit board is mounted inside a Cu box, anchored at the mixing chamber of a dilution refrigerator operating at $T < 30$ mK.

We now discuss the requirements necessary for the readout of the flux qubit. The measurement process disturbs the qubit state, resulting in energy relaxation with a rate $\Gamma_{r}$ and randomization of the phase of the state wavefunction or dephasing with a rate $\Gamma_{\phi}$. For a meaningful measurement the relaxation time $T_{r}=1/\Gamma_{r}$ needs to be much larger than the discrimination time $T_{discr}$, which is the time required to obtain enough information to infer the qubit state. It is defined as
\begin{equation}\label{eq_meastime}
T_{discr}=S_{V}/(V_{1}-V_{0})^{2},
\end{equation}
where $S_{V}$ is the spectral density of the detector output voltage noise and $V_{0}$ and $V_{1}$ are the voltage values corresponding to the qubit in the ground and in the excited state, respectively. We have calculated the decoherence rates for the case of moderate AC current driving. With the details due to be presented separately, the energy relaxation and the dephasing rate are given by
\begin{equation}\label{eq_relaxation}
\Gamma_{r}=\frac{1}{2\hbar^{2}}\sin^{2}\theta~k^{2}(\gamma_{0})[S_{\gamma_{}}^{+}(\omega_{01}+2\pi\nu_{})+S_{\gamma_{}}^{+}(\omega_{01}-2\pi\nu_{})]
\end{equation}
and
\begin{equation}\label{eq_relaxation}
\Gamma_{\phi}=\frac{\Gamma_{r}}{2}+\frac{1}{\hbar^{2}}\cos^{2}\theta~k^{2}(\gamma_{0})[S_{\gamma_{}}^{+}(2\pi\nu_{})].
\end{equation}
Here \(\tan\theta = \Delta /\varepsilon\), \(\omega_{01}=\sqrt{\epsilon^{2}+\Delta^{2}}/\hbar\) and \(S^{+}_{\gamma_{}}\) is the Fourier transform of the symmetrized correlation function of the SQUID phase operator $\gamma_{}$. The coupling of the qubit to the SQUID $k(\gamma_{0})$ is given by
\begin{equation}\label{coupling}
k(\gamma_{0})=MI_{p}I_{circ,SQ}\gamma_{0}
\end{equation}
where \(M\) is the qubit-SQUID mutual inductance and $I_{circ,SQ}$ is the SQUID circulating current, given by \(I_{c0}\sin(\pi\Phi_{SQ}/\Phi_{0})\). The amplitude of the SQUID phase oscillations \(\gamma_{0}\) is proportional to $I_{AC}$ and thus the coupling between the detector and the qubit can be changed in-situ. The discrimination time, the relaxation time and the dephasing time $T_{\phi}=1/\Gamma_{\phi}$ are inversely proportional to $\gamma_{0}^2$. This fact illustrates the tradeoff between obtaining information about the state of a quantum system and the disturbance introduced by the measurement process. In practice, the discrimination time is limited by the noise of the first amplification stage. An analysis of the circuit shown in Fig.~\ref{fig1}c shows that efficient readout is possible if an optimized cryogenic amplifier~\cite{bradley_1999_1} is used for readout. A second requirement for the detector is that there is no significant influence on the qubit during state preparation and manipulation, when no measurement is performed. This case has been discussed in~\cite{vanderwal_2003_1}, and for our circuit this influence is found to be negligible on the time scales necessary for qubit manipulation.

\begin{figure}[!]
\includegraphics[width=3.4 in]{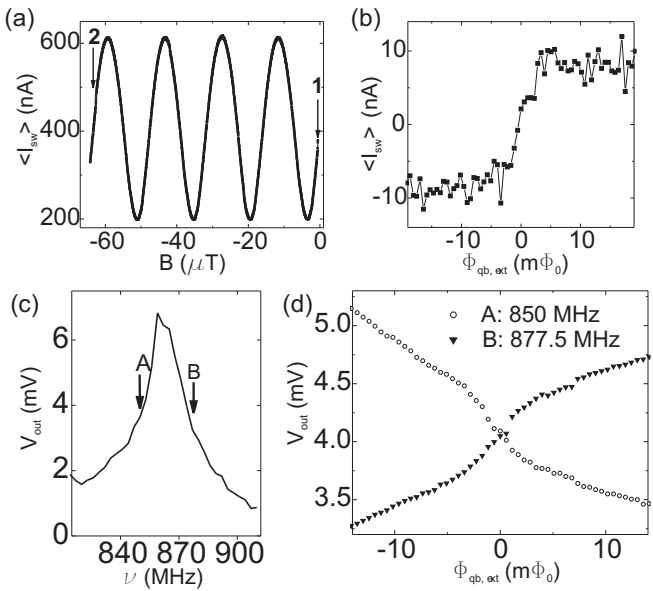}
\caption{\label{fig2} (a) SQUID average switching current versus magnetic field. (b) SQUID average switching current at step 2, with a linear background subtracted. (c) Frequency response of the readout circuit at a magnetic field corresponding to qubit step 2. (d) The response to magnetic flux changes for two operating frequencies, indicated in (c).}
\end{figure}

An initial characterization of the SQUID and qubit was done by measuring the switching current of the DC-SQUID~\cite{vanderwal_2000_1}. The dependence of the average SQUID switching current on the magnetic field, shown in~Fig. \ref{fig2}a, had a period $\Delta B_{SQ} = 16.1$ $\mu$T. At a few equidistant positions separated by $\Delta B = 21$ $\mu$T the SQUID modulation curve shows an increased slope. We attribute these features to the qubit generated flux, which changes sign when $\Phi_{qb}=(2n+1)\Phi_{0}/2$. This is confirmed by the fact that $\Delta B/\Delta B_{SQ} \simeq A_{SQ}/A_{qb}$, indicating a period of 1 $\Phi_{0}$ in the PCQ loop, and the variation of the average switching current is in the direction of increasing flux (see Fig.~\ref{fig1}b). The average switching current of the SQUID around $B = -63.7$ $\mu$T is plotted versus $\Phi_{qb,ext}$ in Fig.~\ref{fig2}b. For the inductive measurements we concentrate on the qubit steps at $B = -0.6$ $\mu$T and at $B = -63.7$ $\mu$T, separated by $3\Phi_{0}$ (indicated as steps 1 and 2).

Now, turning to the inductive method, we measured the amplitude of the SQUID AC voltage versus the frequency of the injected AC current. A resonance peak was found at a frequency $\nu_{0}$, dependent on $\Phi_{SQ,ext}$. The width of this peak and the magnetic field dependence are consistent with the expected properties of the LC resonance for the circuit shown in Fig.~\ref{fig1}c. From a fit of the dependence $\nu_{0}(\Phi_{SQ,ext})$ using the relation $\nu_{0}(\Phi_{SQ,ext})=1/(2\pi\sqrt{(L_{s}+L_{J}(\Phi_{SQ,ext}))C})$, we find $L_{s} \simeq 2.35$ nH and the minimum value of the Josephson inductance $L_{J, min} \simeq 0.31$ nH. The obtained $L_{J, min}$ is in agreement with the value of the maximum critical current of 1.0 $\mu$A, as calculated using the value of the SQUID normal state resistance.

Figure~\ref{fig2}c shows the circuit resonance peak at a magnetic field corresponding to the qubit step 2. Here, the circuit resonance frequency $\nu_{0}$ increases with increasing magnetic field. After fixing a SQUID driving frequency either smaller or larger than the resonance frequency, we vary the external magnetic field (see Fig.~\ref{fig2}d). Superimposed on the smooth slope, the qubit step is clearly visible, similar to the data presented in Fig.~\ref{fig2}b.

\begin{figure}[!]
\includegraphics[width=3.4 in]{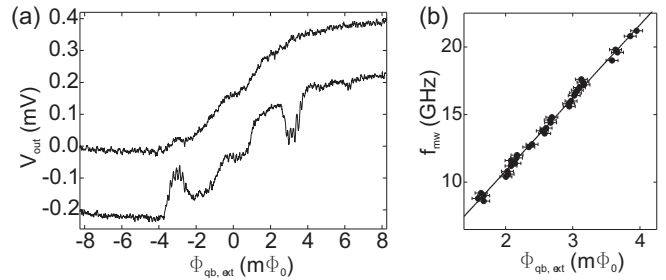}
\caption{\label{fig3} (a) Spectroscopy measurements at $f_{mw} = 17$ GHz, for $T_{mw} = 1$ $\mu$s, $T_{d} = 0$ $\mu$s, and $T_{m} = 2$ $\mu$s, for $I_{mw} = 0$ (top curve) and for $I_{mw} \sim 7$ $\mu$A (bottom curve). The readout circuit has $\nu_{0} = $ 840 MHz and is operated at $\nu = 855$ MHz and $\gamma_{0} \simeq 0.5$. The data is plotted after averaging over flux intervals of $49$ $\mu\Phi_{0}$ and $42$ $\mu\Phi_{0}$, respectively, and the two curves are offset. (b) $f_{mw}$ vs flux in the PCQ loop (dots) and a linear fit through the origin (continuous line).}
\end{figure}

As a next step we performed spectroscopy and measured the qubit relaxation time. For this purpose we first apply a microwave burst of frequency $f_{mw}$ and duration $T_{mw}$ to excite the qubit. After a delay time $T_{d}$ the SQUID detector is switched on by injecting the current $I_{AC}$ and the output voltage is measured for a time $T_{m}$. The time $T_{m}$ is taken larger than the SQUID response time, $Q/(2\pi\nu_{0}) \sim 8$ ns. To improve statistics, we average over typically 1000 of these measurement sequences.

Figure~\ref{fig3} shows the spectroscopy measurement results. In Fig.~\ref{fig3}a, $V_{out}$ is plotted against magnetic flux at step 1, after substraction of the linear background. If microwaves are applied, a peak and a dip are observed at positions symmetrically around the center of the step, with a height equal to almost half the step size. The peak and the dip correspond to the condition that the frequency of the applied MW burst matches the qubit energy level separation, which results in an incoherent mixture of the ground and excited states. The amplitude of the peaks and dips decreases with decreasing microwave power (data not shown) and the position of the peaks depends on $f_{mw}$, as shown in Fig.~\ref{fig3}b. We did not reliably identify peaks and dips for $f_{mw} < 8.5$ GHz and therefore we could not determine $\Delta$. A linear fit of the peak positions gives a value of the maximum persistent current $I_{p} = 870 \pm 30$ nA; this agrees well with the value estimated from the amplitude of the qubit step, and is in reasonable agreement with the design value of 660 nA. We have performed the same measurements at qubit step 2 and we obtained similar results.

\begin{figure}[!]
\includegraphics[width=3.0 in]{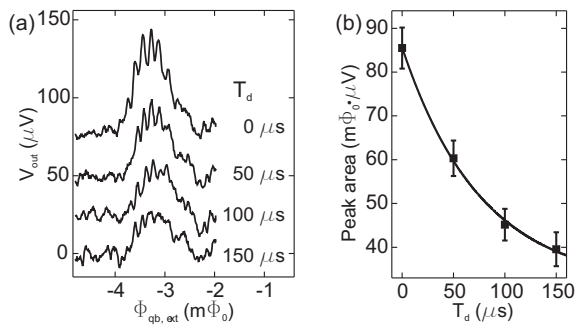}
\caption{\label{fig4}(a) Spectroscopy peak at 17 GHz (plotted data is averaged over $\Delta\Phi = 56.2$ $\mu\Phi_{0}$). (b) Spectroscopy peak area vs delay time and a fit with an exponential decay dependence.}
\end{figure}

Both the spectroscopy peaks and dips display a weak variation, periodic with the applied magnetic flux (see Fig.\ref{fig4}a). This period is the same in the peak and the dip and independent of microwave frequency and power. The position of the individual subpeaks changes linearly with $f_{mw}$, following the peaks and dips in which they are contained. In terms of qubit energies, the sub-peaks splitting $\Delta\Phi_{s}$ corresponds to $I_{p}\Delta \Phi_{s}/h = 880 \pm 130$ MHz,which matches the SQUID resonance frequency $\nu$ within the experimental uncertainty. This suggests that the observed substructure originates from transitions between the energy levels of the coupled qubit-SQUID system. The coherence time of the qubit, estimated from the width of the spectroscopy subpeaks, is $T_{2} \simeq 1$ ns. This short decoherence time is not due to the measurement, as the SQUID AC driving is switched off during qubit excitation. The curves in Fig.~\ref{fig3}a show a region, in the center of the step, characterized by a smaller slope. The width of this region decreases with decreasing amplitude of the SQUID AC driving. We attribute it to the influence of the AC magnetic field generated by the measuring SQUID on the qubit.

In order to determine the relaxation time, we have measured the spectroscopy peak using different time delays $T_{d}$. Figure~\ref{fig4} shows the results obtained at qubit step 1 for a level separation of 17 GHz. The area of the Lorentzian curves fitting the peaks is plotted versus $T_{d}$ in Fig.~\ref{fig4}b. A fit with an exponential decay gives the value of the relaxation time ${T_{1} = 77 \pm 12}$ $\mu$s. The obtained value was the same for peak or dip and independent of the applied microwave power.

From the measurements we estimate the parameters characterizing the readout efficiency, referred to the end of the amplification chain: \(\sqrt{S_{V}}\simeq 4 \) \(\mu V/\sqrt{Hz}\) and $V_{1}-V_{0} \simeq 800$ $\mu V$, which implies $T_{r}/T_{discr}\simeq 3.1$. The measurement fidelity, defined as the probability to infer the qubit state correctly, based on the detector output voltage, is  $\sim $70\%. Further improvement of the efficiency is possible by optimizing the SQUID circuitry. Using an on-chip circuit with small stray inductances and optimized power transfer to the amplifier can reduce the measurement time by a factor of 100. Also, the present temperature $T_n\sim20$ K of our amplifier, can be reduced to $T_{n}\sim3$ K by using an optimized cryogenic amplifier or to $T_{n}<100$ mK by using a SQUID amplifier~\cite{muck_2001_1}, with proportional decrease in the measurement time.

The preliminary experiment of this paper, spectroscopy measurements on a flux qubit, yielded the proof of principle for this method. The decoherence time was short, but feasible changes of the circuit can remedy this. The  absence of the spurious back-action effects associated with switching of Josephson junctions, the continuous nature of the detection and the in-situ tunability of the qubit-detector coupling allow for fundamental experimental studies of the quantum measurement process and for correlation measurements on two qubits.

We thank Alexander ter Haar and Hannes Majer for discussions. This work was supported by the Dutch Organization for Fundamental Research on Matter (FOM), the European Union SQUBIT project, and the U.S. Army Research Office (grant DAAD 19-00-1-0548).


\end{document}